%% file: DESY-04-070.tex
\documentclass[zpreprint,zbstnp]{zeus_paper}
\usepackage[english]{babel}
\input ./zeus_def.tex

\def\P{{\rm I\kern-.15em P}}
\def\rel{$p_{T}^{\rm{rel}}$}
\begin{document}
\include{paper-tit}
\include{paper-aut}
\include{paper-txt}
\include{paper-ref}
\include{paper-tab}
\include{paper-fig}
%
%
\end{document}

%% file: zeus_def.tex
\chardef\usc=95
\chardef\til=126
\catcode`\@=11 
\DeclareRobustCommand\xdotspace{\futurelet\@let@token\@xdotspace}
\def\@xdotspace{%
  \ifx\@let@token.\else
  \ifx\@let@token\bgroup.\else
  \ifx\@let@token\egroup.\else
  \ifx\@let@token\/.\else
  \ifx\@let@token\ .\else
  \ifx\@let@token~.\else
  \ifx\@let@token!.\else
  \ifx\@let@token,.\else
  \ifx\@let@token:.\else
  \ifx\@let@token;.\else
  \ifx\@let@token?.\else
  \ifx\@let@token/.\else
  \ifx\@let@token'.\else
  \ifx\@let@token).\else
  \ifx\@let@token-.\else
  \ifx\@let@token\@xobeysp.\else
  \ifx\@let@token\space.\else
  \ifx\@let@token\@sptoken.\else
   .\space
   \fi\fi\fi\fi\fi\fi\fi\fi\fi\fi\fi\fi\fi\fi\fi\fi\fi\fi}
\catcode`\@=12 

\newcommand{\stru}[2]{%
   \relax\ifmmode\hbox{\vrule height#1 depth#2 width0pt}%
   \else\vrule height#1 depth#2 width0pt\fi}

\newcommand{\Ronum}[1]{\uppercase\expandafter{\romannumeral#1}}
\newcommand{\ronum}[1]{\expandafter{\romannumeral#1}}
\DeclareRobustCommand{\LaTeXZ}{%
  \LaTeX\kern-.05em4\kern-.1em
  {\raisebox{-0.2ex}{$\scriptstyle\text{ZEUS}$}}\xspace}

\newcommand{\fig}[1]{Fig.~\ref{fig-#1}}
\newcommand{\Fig}[1]{Figure~\ref{fig-#1}}

\newcommand{\tab}[1]{Table~\ref{tab-#1}}

\newcommand{\Sect}[1]{Section~\ref{sec-#1}}

\DeclareMathAlphabet{\mathbf}{OT1}{cmr}{bx}{sl}
\newcommand{\eVdist}{\kern-0.06667em}

\newcommand{\Gev}{{\text{Ge}\eVdist\text{V\/}}}

\newcommand{\gev}{{\,\text{Ge}\eVdist\text{V\/}}}

\newcommand{\pb}{\,\text{pb}}

\newcommand{\pbi}{\,\text{pb}^{-1}}

\newcommand{\Tesla}{\,\text{T}}

\newcommand{\slashfrac}[2]{%
  \raisebox{0.5ex}{\ensuremath #1}\kern-0.12em/\kern-0.08em
  \raisebox{-.8ex}{\ensuremath #2}}

\newcommand{\sqr}[3]{%
    {\vcenter{\hrule height.#3ex\hbox{\vrule width.#2ex height#1ex
     \kern#1ex\vrule width.#3ex}\hrule height.#2ex}}}

\newcommand{\widebar}[1]{%
   \mkern1.5mu\overline{\mkern-1.5mu#1\mkern-1.mu}\mkern1.mu}
\catcode`\@=11 
\newcommand{\parenbar}{\mathpalette\p@renb@r}
\def\p@renb@r#1#2{\vbox{%
  \ifx#1\scriptscriptstyle \dimen@.7em\dimen@ii.2em\else
  \ifx#1\scriptstyle \dimen@.8em\dimen@ii.25em\else
  \dimen@1em\dimen@ii.4em\fi\fi \offinterlineskip
  \ialign{\hfill##\hfill\cr
    \vbox{\hrule width\dimen@ii}\cr
    \noalign{\vskip-.3ex}%
    \hbox to\dimen@{$\mathchar300\hfil\mathchar301$}\cr
    \noalign{\vskip-.3ex}%
    $#1#2$\cr}}}
\catcode`\@=12 

\newcommand{\bbar}{\widebar{b}}

\newcommand{\als}{\alpha_s}

\newcommand{\IP}{{\rm I$\kern-0.01667em$P}\xspace}

\mathchardef\qsm=63
\mathchardef\pls=43
\mathchardef\mns=512
\mathchardef\plm=518
\mathchardef\eql=61
\mathchardef\smallleft=300
\mathchardef\smallright=301
\mathchardef\les=316
\mathchardef\gre=318
\mathchardef\leq=532
\mathchardef\grq=533

\catcode`\@=11 
\newcounter{pict@width}
\newcounter{pict@height}
\newlength{\pict@scale}
\newcommand{\psfigadd}[4]{%
\setcounter{pict@width}{1*\ratio{#2+\pict@scale/2}{\pict@scale}}
\setcounter{pict@height}{1*\ratio{#3+\pict@scale/2}{\pict@scale}}
\setlength{\unitlength}{\pict@scale}
\hbox to #2{\hspace{-\fill}\begin{picture}(\thepict@width,\thepict@height)
\put(0,0){\psfig{figure=#1,width=#2,height=#3,clip=}}
\SetScale{0.283466457}
\SetWidth{1.763889}
{#4}
\end{picture}}
}
\newcounter{pict@widthfst}
\newcounter{pict@widthscd}
\newcounter{pict@widthtot}
\newcommand{\psfigaddtwo}[7]{%
\setcounter{pict@widthfst}{1*\ratio{#2+\pict@scale/2}{\pict@scale}}
\setcounter{pict@widthscd}{1*\ratio{#2+#4+\pict@scale/2}{\pict@scale}}
\setcounter{pict@widthtot}{1*\ratio{#2+#4+#6+\pict@scale/2}{\pict@scale}}
\setcounter{pict@height}{1*\ratio{#3+\pict@scale/2}{\pict@scale}}
\setlength{\unitlength}{\pict@scale}
\hbox{\hspace{-\fill}\begin{picture}(\thepict@widthtot,\thepict@height)
\put(0,0){\psfig{figure=#1,width=#2,height=#3,clip=}}
\put(\thepict@widthscd,0){\psfig{figure=#5,width=#6,height=#3,clip=}}
\SetScale{0.283466457}
\SetWidth{1.763889}
{#7}
\end{picture}}
}
\newcommand{\psfigror}[4]{%
\setcounter{pict@width}{1*\ratio{#2+\pict@scale/2}{\pict@scale}}
\setcounter{pict@height}{1*\ratio{#3+\pict@scale/2}{\pict@scale}}
\setlength{\unitlength}{\pict@scale}
\hbox{\begin{picture}(\thepict@width,\thepict@height)
\put(0,\thepict@height){\psfig{figure=#1,width=#3,height=#2,clip=,angle=270}}
\SetScale{0.283466457}
\SetWidth{1.763889}
{#4}
\end{picture}}
}
\newcommand{\psfigrol}[4]{%
\setcounter{pict@width}{1*\ratio{#2+\pict@scale/2}{\pict@scale}}
\setcounter{pict@height}{1*\ratio{#3+\pict@scale/2}{\pict@scale}}
\setlength{\unitlength}{\pict@scale}
\hbox{\begin{picture}(\thepict@width,\thepict@height)
\put(0,0){\psfig{figure=#1,width=#3,height=#2,clip=,angle=90}}
\SetScale{0.283466457}
\SetWidth{1.763889}
{#4}
\end{picture}}
}
\catcode`\@=12 
\newlength\listtextwidth

\catcode`\@=11 
\newlength{\@tabfninsert}
\newlength{\@tabfnwidth}
\newcommand{\tabfootnote}[2]{%
  \setlength{\@tabfninsert}{0.8em}
  \setlength{\@tabfnwidth}{\textwidth}
  \addtolength{\@tabfnwidth}{-\@tabfninsert}
  \addtolength{\@tabfnwidth}{-0.4em}
  \noindent\makebox[\@tabfninsert][r]{\footnotesize$^{#1}$\hfil}\hfill%
  \parbox[t]{\@tabfnwidth}{\footnotesize #2\hfill}}
\catcode`\@=12 

%% file: paper-tit.tex
\prepnum{DESY--04--070}
%
%
\title{
Measurement of beauty production \\
in deep inelastic scattering at HERA 
}                                                       
\author{ZEUS Collaboration}
\date{May 2004}
\abstract{
The beauty production cross section for deep inelastic scattering events 
with at least one hard jet in the Breit frame together with a muon 
has been measured,
for photon virtualities $Q^2>2$ GeV$^2$, with the ZEUS
detector at HERA using integrated luminosity of $72\pbi$.  The total
visible cross section is $\sigma_{b\bbar}(ep
\rightarrow e~{\rm jet}~\mu~X) =
40.9 \pm 5.7\;{\rm (stat.)} ^{+6.0}_{-4.4} {\rm (syst.)} \pb.$ The
next-to-leading order QCD prediction lies about 2.5 standard
deviations below the data.  The differential cross sections are in
general consistent with the NLO QCD predictions; however at low values
of $Q^2$, Bjorken $x$, and muon transverse momentum, and high values of
jet transverse energy and muon pseudorapidity, the prediction is about
two standard deviations below the data.}
\makezeustitle

%% file: paper-aut.tex
\def\3{\ss}                                                                                        
\newcommand{\address}{ }                                                                           
\pagenumbering{Roman}                                                                              
                                                   %
\begin{center}                                                                                     
{                      \Large  The ZEUS Collaboration              }                               
\end{center}                                                                                       
  S.~Chekanov,                                                                                     
  M.~Derrick,                                                                                      
  J.H.~Loizides$^{   1}$,                                                                          
  S.~Magill,                                                                                       
  S.~Miglioranzi$^{   1}$,                                                                         
  B.~Musgrave,                                                                                     
  \mbox{J.~Repond},                                                                                
  R.~Yoshida\\                                                                                     
 {\it Argonne National Laboratory, Argonne, Illinois 60439-4815}, USA~$^{n}$                       
\par \filbreak                                                                                     
  M.C.K.~Mattingly \\                                                                              
 {\it Andrews University, Berrien Springs, Michigan 49104-0380}, USA                               
\par \filbreak                                                                                     
  N.~Pavel \\                                                                                      
  {\it Institut f\"ur Physik der Humboldt-Universit\"at zu Berlin,                                 
           Berlin, Germany}
% ...
\newpage                                                                                           
$^{\    1}$ also affiliated with University College London, UK \\                                  
% ...
                                                           %
                                                           %
\newpage   
                                                           %
                                                           %
\begin{tabular}[h]{rp{14cm}}                                                                       
$^{a}$ &  supported by the Natural Sciences and Engineering Research Council of Canada (NSERC) \\  
$^{b}$ &  supported by the German Federal Ministry for Education and Research (BMBF), under        
          contract numbers HZ1GUA 2, HZ1GUB 0, HZ1PDA 5, HZ1VFA 5\\                                
$^{c}$ &  supported in part by the MINERVA Gesellschaft f\"ur Forschung GmbH, the Israel Science   
          Foundation (grant no. 293/02-11.2), the U.S.-Israel Binational Science Foundation and    
          the Benozyio Center for High Energy Physics\\                                            
$^{d}$ &  supported by the German-Israeli Foundation and the Israel Science Foundation\\           
$^{e}$ &  supported by the Italian National Institute for Nuclear Physics (INFN) \\                
$^{f}$ &  supported by the Japanese Ministry of Education, Culture, Sports, Science and Technology 
          (MEXT) and its grants for Scientific Research\\                                          
$^{g}$ &  supported by the Korean Ministry of Education and Korea Science and Engineering          
          Foundation\\                                                                             
$^{h}$ &  supported by the Netherlands Foundation for Research on Matter (FOM)\\                   
$^{i}$ &  supported by the Polish State Committee for Scientific Research, grant no.               
          620/E-77/SPB/DESY/P-03/DZ 117/2003-2005\\                                                
$^{j}$ &  partially supported by the German Federal Ministry for Education and Research (BMBF)\\   
$^{k}$ &  supported by RF President grant N 1685.2003.2 for the leading scientific schools and by  
          the Russian Ministry of Industry, Science and Technology through its grant for           
          Scientific Research on High Energy Physics\\                                             
$^{l}$ &  supported by the Spanish Ministry of Education and Science through funds provided by     
          CICYT\\                                                                                  
$^{m}$ &  supported by the Particle Physics and Astronomy Research Council, UK\\                   
$^{n}$ &  supported by the US Department of Energy\\                                               
$^{o}$ &  supported by the US National Science Foundation\\                                        
$^{p}$ &  supported by the Polish Ministry of Scientific Research and Information Technology,      
          grant no. 112/E-356/SPUB/DESY/P-03/DZ 116/2003-2005\\                                    
$^{q}$ &  supported by the Polish State Committee for Scientific Research, grant no.               
          115/E-343/SPUB-M/DESY/P-03/DZ 121/2001-2002, 2 P03B 07022\\                              
\end{tabular}                                                                                      
                                                           %
                                                           %

%% file: paper-txt.tex
\pagenumbering{arabic} 
\pagestyle{plain}







%
\input{txt-int}
%
\input{txt-ana}

%
\input{txt-res}
%
\input{txt-con}

%
\input{txt-ack}
\vfill\eject

%% file: txt-int.tex
\section{Introduction}
\label{sec-int}
Deep inelastic scattering (DIS) offers a unique opportunity to study the production
mechanism of bottom ($b$) quarks via the strong interaction
in a clean environment where a point-like 
projectile, a photon with a virtuality $Q^2$, collides with a proton. 
Due to the large centre-of-mass energy,
$b\bar{b}$ pairs are copiously produced at the electron-proton
collider HERA. 
The large $b$-quark mass provides a hard scale, making perturbative 
Quantum Chromodynamics (QCD) applicable.
However, a hard scale can also be given by the transverse
jet energy and by $Q$.
The presence of two or more scales can lead to large logarithms in the
calculation which can possibly spoil the convergence of the perturbative
expansion. Precise differential cross-section measurements are therefore
needed to test the theoretical understanding of $b$-quark production
in strong interactions.

The cross sections for $b$-quark production in strong interactions have
been measured 
in proton-antiproton collisions at the S${\rm p\bar{p}}$S~\cite{pl:b256:121,*pl:b262:497} and
the Tevatron~\cite{prl:71:500,*prl:71:2396,*pr:d53:1051,*pr:d55:2546,
*pl:b487:264} and, more recently, in two-photon 
interactions at LEP~\cite{pl:b503:10} and
in $\gamma p$ interactions at HERA~\cite{pl:b467:156,epj:c18:625,*zeus:php}.
Some of the $b$-production
cross sections are significantly above
the QCD expectations calculated to next-to-leading order (NLO)
in the strong coupling constant, $\als$. 

This paper reports the first measurement of $b$-quark production in DIS at HERA, 
in the reaction with at least one hard jet in the Breit 
frame~\cite{feynman:1972:p-hi} and a muon, from a $b$ decay, in the final state:
$$ep \rightarrow e~b~\bar{b}~X \rightarrow e + {\rm jet} + \mu + X.$$
In the Breit frame, defined by ${\boldsymbol \gamma}+2x\mathbf{P} = 0,$ where 
${\boldsymbol \gamma}$ is the momentum of the exchanged photon, $x$ is 
the Bjorken scaling variable and 
$\mathbf{P}$ is the proton momentum, a space-like
 photon and a proton collide head-on. 
In this frame, any final-state particle with a high transverse momentum is 
produced by a hard QCD interaction.

In this paper, a measurement of the visible cross section, $\sigma_{b\bbar}$,
is presented, as well as several differential cross sections.
The measured cross sections are compared to  
Monte Carlo (MC) models which use leading order (LO) matrix elements, with the 
inclusion of initial- and final-state parton showers, as well as to NLO QCD calculations. 
All cross sections are measured in a kinematic 
region in which the scattered electron, 
the muon and the jet are well reconstructed in the ZEUS detector.


%% file: txt-ana.tex
\section{Experimental conditions}
\label{sec-ana}

The data used in this measurement were collected during the 1999-2000 HERA 
running period, where a proton beam of $920\gev$ collided with 
a positron or electron beam of 
$27.5\gev$, corresponding to a centre-of-mass energy of 318 \gev.
The total integrated luminosity was ($72.4 \pm 1.6) \pbi.$

A detailed description of the ZEUS detector can be found 
elsewhere~\cite{pl:b293:465,zeus:1993:bluebook}.
A brief outline of the components that are most relevant for this analysis 
is given below.
The high-resolution uranium--scintillator calorimeter
(CAL)~\cite{nim:a309:77,*nim:a309:101,*nim:a321:356,*nim:a336:23} consists
of three parts: the forward (FCAL), the barrel (BCAL) and the rear (RCAL)
calorimeters. Each part is subdivided transversely into towers and
longitudinally into one electromagnetic section (EMC) and either one
(in RCAL) or two (in BCAL and FCAL) hadronic sections (HAC). The
smallest subdivision of the calorimeter is called a cell. 
The CAL energy resolutions, as measured under test-beam conditions, 
are $\sigma(E)/E=0.18/\sqrt{E\,({\rm GeV})}$ for
electrons and $\sigma(E)/E=0.35/\sqrt{E\,({\rm GeV})}$ for hadrons.

Charged particles are tracked in the central tracking detector
(CTD)~\cite{nim:a279:290,*npps:b32:181,*nim:a338:254}, which operates in a
magnetic field of $1.43\Tesla$ provided by a thin superconducting
solenoid. The CTD consists of 72~cylindrical drift-chamber
layers, organised in nine superlayers covering the
polar-angle\footnote{The ZEUS coordinate system is a right-handed
  Cartesian system, with the $Z$ axis pointing in the proton beam
  direction, referred to as the ``forward direction'', and the $X$
  axis pointing left towards the centre of HERA. The coordinate origin
  is at the nominal interaction point.}
region \mbox{$15^\circ<\theta<164^\circ$}. The transverse-momentum
resolution for full-length tracks can be parameterised as
$\sigma(p_T)/p_T=0.0058p_T\oplus0.0065\oplus0.0014/p_T$, with $p_T$ in
$\Gev$. 

The position of electrons\footnote{
Hereafter ``electron'' refers both to electrons and positron.}
scattered at small angles to the electron beam
direction was measured using the small-angle rear tracking detector
(SRTD)~\cite{nim:a401:63, epj:c21:443}.
The SRTD is attached to the front face of
the RCAL and consists of two planes of scintillator strips, 1~cm wide and
0.5~cm thick, arranged orthogonally.

The muon system consists of tracking detectors (forward, barrel and rear muon chambers:
FMUON~\cite{zeus:1993:bluebook}, B/RMUON~\cite{nim:a333:342}), which are placed inside and
outside a magnetised iron yoke surrounding the CAL and cover polar angles from 10$^\circ$
to 171$^\circ$.
The barrel and rear inner muon chambers cover polar angles from 34$^\circ$ to 135$^\circ$.

The luminosity was measured from the rate of the bremsstrahlung process
$ep\rightarrow e \gamma p$. The resulting small-angle energetic photons
were measured by the luminosity
monitor~\cite{desy-92-066,*zfp:c63:391,*acpp:b32:2025}, a
lead-scintillator calorimeter placed in the HERA tunnel at $Z=-107$ m.

\section{Event Selection}
\label{sec-es}

Events were selected online via a three-level trigger 
system~\cite{zeus:1993:bluebook,proc:chep:1992:222}.
The trigger required a localised energy deposit in the EMC
consistent with that of a scattered electron. 
At the third level, where a full event reconstruction is available,
a muon was required, defined by a track in the CTD loosely matching a track segment 
in the inner part of the B/RMUON chambers.

The scattered electron candidate was identified from the pattern of energy deposits in 
the CAL~\cite{nim:a365:508,*nim:a391:360}. The energy ($E_e$) and polar angle ($\theta_e$) 
of the electron are measured by combining the impact position at the calorimeter with the
event vertex. The impact position is measured from the calorimeter cells
associated with the electron candidate, but the CTD ($\theta_e <
157^\circ$) and SRTD ($\theta_e > 162^\circ$) detectors are
used to improve the measurement whenever the electron trajectory lies
within the respective regions of acceptance. 

The reconstruction of $Q^2$ was based on the measurement of the scattered 
electron energy and polar angle~\cite{proc:hera:1991:23}. 
The Bjorken scaling variables $x$ 
and $y$ were reconstructed using the $\Sigma$-method, which allows the 
determination of the estimator $y_{\Sigma}$ independently of initial state photon radiation by 
reconstructing the incident electron energy~\cite{nim:a361:197}.

Events were selected \cite{thesis:chiochia:2003} by requiring the 
presence of at least one muon in the final state and at least one jet 
in the Breit frame. 
The final sample was selected in four steps:
\begin{enumerate}
\item Inclusive DIS event selection:
  \begin{itemize}
  \item a well reconstructed scattered electron was required with energy greater 
    than $10\gev$, $Q^2>2$ GeV$^2$, $y_{JB}> 0.05$ and $y_{\Sigma}<0.7$, 
    where $y_{JB}$ is the $y$ variable reconstructed using the Jacquet-Blondel 
    method~\cite{proc:epfacility:1979:391};
  \item for events with the scattered electron reconstructed within
    the SRTD acceptance the impact position of the 
    electron was required to be outside a box defined by $|X_e|<12$ cm and $|Y_e|<6$ cm.
    For events without SRTD information, a box cut on the face of the
    RCAL of $|X_e|<12$ cm and $|Y_e|<10$ cm was used.
    This cut removed electron candidates  near the inner edge of the
    RCAL beampipe hole;
  \item to reduce the background from collisions of real photons with 
    protons (photoproduction), where the scattered electron escapes down the rear beampipe, 
    the variable $E-p_Z$ was required to be in the range 
    \mbox{$40 < E-p_Z < 65$ \gev}. The variable $E-p_Z$ was defined as the 
    difference of the total energy and the longitudinal component of the total momentum,
    calculated using final-state objects, reconstructed from tracks and energy deposits
    in the calorimeter;                 
   \item the reconstructed event vertex was required to 
    lie within 50 cm of the nominal interaction point.
  \end{itemize}

\item Muon finding:

Muons were identified by requiring a track segment in 
both the inner and outer parts of the BMUON or RMUON chambers. 
The reconstructed muons were matched in space and momentum with a track found in the CTD, 
with a $\chi^2$ probability greater than 1\%.
This cut rejected the background from muons coming from $K^{\pm}$ and $\pi^{\pm}$ 
decays and from particles produced in hadronic showers in the CAL that may 
be misidentified as muons. 
In addition, cuts on the muon momentum, $p^{\mu}$,
the muon transverse momentum, $p_T^{\mu}$ and the muon
pseudorapidity, $\eta^{\mu}$, were applied:
\begin{itemize}
\item $-0.9 < \eta^{\mu} < 1.3$ and $p_T^{\mu}>2\gev$ corresponding to the BMUON region;
\item $-1.6 < \eta^{\mu} < -0.9$ and $p^{\mu}>2\gev$  corresponding to the RMUON region.
\end{itemize}
The reconstruction efficiency of the muon chambers was calculated 
separately for BMUON and RMUON using an independent data sample of di-muon 
events produced in photon-proton collisions~\cite{thesis:turcato:2003}.
This data sample consisted of elastic and quasi-elastic Bethe-Heitler events
($\gamma \gamma \rightarrow \mu^+ \mu^-$) and $J/\psi$ production.

The data sample was selected from events triggered by the inner
  muon chambers. Two tracks, reconstructed in the CTD,
  with transverse momentum greater than 1 GeV and associated
  with energy deposits in the CAL consistent with a minimum-ionising
  particle were required.
One of the CTD tracks was required
to point to the muon chamber that triggered the event, and the other was used
to measure the muon efficiency, defined as the ratio
of the number of tracks satisfying the muon matching requirement to the
total number of tracks.
The measured muon-reconstruction efficiencies are between 20\% and 40\%, 
depending on the region of the muon chambers and on the muon transverse
momentum.
\item Jet finding:

Hadronic final-state objects were boosted to the Breit frame and clustered 
into jets
using the $k_T$ cluster algorithm ({\sc Ktclus})~\cite{np:b406:187} in its 
longitudinally invariant inclusive mode~\cite{pr:d48:3160}. 
The four-momenta of the hadronic final-state objects were calculated from the measured
energy and angle, assuming the objects to be massless.
The $p_T$ recombination scheme was used. 
Reconstructed muons were included in the clustering procedure.  
Events were required to have at least one jet with transverse energy measured
in the Breit frame, $E_{T,\rm{jet}}^{\rm{Breit}}$, above $6\gev$ and 
within the detector \mbox{acceptance}, $-2 < \eta_{\rm{jet}}^{\rm{lab}} < 2.5$, 
where $\eta_{\rm{jet}}^{\rm{lab}}$ is the jet
pseudorapidity in the laboratory frame.
\item Muon-jet association:

The muons in the sample were associated with the jet containing
the corresponding hadronic final-state object using the 
{\sc Ktclus} information. The associated jet was not necessarily the 
jet satisfying the jet requirements above. To ensure that the associated jet was well 
reconstructed, it was required to have $E_{T,\rm{jet}}^{\rm{Breit}}>4\gev$.

\end{enumerate}

After these selection cuts, 941 events remained.

\section{Monte Carlo simulation and NLO QCD calculations}
\label{sec-mc-theo}

To correct the results for detector effects and to extract the 
fraction of events from $b$ decays, two MC simulations were used: 
{\sc Rapgap 2.08/06} as default and 
{\sc Cascade 1.00/09} for systematic checks.
The predictions of the MC simulations were also compared to the final results.

The program {\sc Rapgap 2.08/06}~\cite{cpc:86:147} is an event generator 
based on leading-order (LO) matrix elements, with 
higher-order QCD radiation simulated in the leading-logarithmic approximation 
using initial- and final-state parton showers based on the DGLAP equations
\cite{sovjnp:15:438,*sovjnp:20:94,*jetp:46:641,*np:b126:298}.
To estimate the background, samples with light and charm
quarks in the final state were produced. The process in which a $b\bar{b}$ pair 
is produced in photon-gluon fusion was used to simulate the signal.
The charm and $b$-quark masses were set to 1.5\gev and 5\gev, 
respectively. The CTEQ5L~\cite{epj:c12:375} parameterisation of the
proton parton densities was used. Heavy-quark hadronisation 
was modelled by the Bowler fragmentation function~\cite{zfp:c11:169}.
The rest of the hadronisation was simulated using the Lund string model~\cite{prep:97:31} 
as implemented in {\sc Jetset 7.4}~\cite{cpc:82:74}. The {\sc Rapgap} MC includes
the LO electroweak corrections calculated using 
{\sc Heracles 4.6.1}~\cite{cpc:69:155}.

The {\sc Cascade 1.00/09} MC~\cite{epj:c19:351,*cpc:143:100} uses the
$O(\alpha_s)$ matrix elements, where the incoming partons can be
off-shell. The parton evolution is based on the CCFM
equations~\cite{np:b296:49,*pl:b234:339,*np:b445:45}, which are derived 
from the principles of $k_T$ factorisation and colour coherence. 
The mass of the $b$ quark was set to 4.75 GeV.

The NLO QCD predictions were evaluated using the {\sc Hvqdis} 
program~\cite{pr:d57:2806,np:b452:109,*pl:b353:535}, which includes only 
point-like photon contributions.
The fragmentation of a $b$ quark into a $B$ hadron was modelled by the
Kartvelishvili function~\cite{pl:b78:615}.
The parameter $\alpha$ was set to 27.5, as obtained by
an analysis~\cite{prl:89:122003} of $e^+e^-$ data~\cite{pl:b512:30}.
The semi-leptonic decay of $B$ hadrons into muons was modelled 
using a parameterisation of the muon momentum spectrum extracted from {\sc Rapgap}.
This spectrum corresponds to a mixture of direct $(b \rightarrow \mu)$ 
and indirect $(b \rightarrow c \rightarrow \mu)$ $B$-hadron decays.
Jets were reconstructed by running the inclusive  $k_T$  algorithm,
   using the  $p_T$ recombination scheme, on the four-momentum of
   the two or three partons generated by the program.
The $b$-quark mass was set to $m_b = 4.75 \gev$ and the renormalisation and 
factorisation scales to $\mu = \sqrt{p_{T,b}^2+m_b^2}$, where 
$p_{T,b}$ is the mean transverse momentum of the $b$ and $\bar{b}$ quarks. 
The CTEQ5F4 proton parton densities~\cite{epj:c12:375} were used. 
The sum of the branching ratios of 
direct  and indirect  decays of $B$ hadrons into muons was fixed to 
the {\sc Jetset 7.4} value of 0.22.

The NLO QCD predictions were multiplied by hadronisation corrections
to compare them to the measured cross sections. The hadronisation 
corrections are defined as the ratio of the cross sections obtained
by applying the jet finder to
the four-momenta of all hadrons, assumed to be massless, 
and that from applying it to the four-momenta of all partons. 
They were evaluated using the {\sc Rapgap} program; they lower the NLO QCD
prediction by typically 10\%.

The uncertainty of the NLO prediction was estimated by varying 
the factorisation and renormalisation scales, $\mu$, by a factor of 2 and 
the $b$-quark mass, $m_b$, between 4.5 and \mbox{5.0 \gev} and adding the 
respective contributions in quadrature. 
Additional uncertainties due to different scale choices and to different 
fragmentation functions are within the quoted uncertainties.
More details of the NLO QCD calculation and of the determination of its
uncertainties can be found elsewhere~\cite{pr:d57:2806,np:b452:109,jhep:09:070}.
%
\section{Extraction of the beauty fraction}
\label{sec-bf}

A significant background to the process under study 
is due to  muons from in-flight decays of pions and kaons. 
Such decay muons are mostly characterised by low momenta and, therefore, 
partly rejected by the cuts $p^\mu > 2$ GeV and $p_T^\mu > 2$  GeV.
In addition, the signal reconstructed in the muon chambers can be due 
to  kaons or pions passing through the CAL.
Muons can also originate from the semi-leptonic decay of charmed hadrons.
These decays produce events topologically similar to those under study.

Due to the large $b$-quark mass, muons from semi-leptonic $b$ decays usually 
have high values of the transverse momentum, \rel, 
 with respect to the axis of the closest jet.
For muons from charm decays and in events induced by light 
quarks, 
the \rel~values are low. 
Therefore, the fraction of events from $b$ decays in the data sample can be 
extracted on a statistical basis by fitting the relative contributions of the 
simulated bottom, charm and light-quark decays to the measured \rel~distribution.

The extraction of the fraction of $b$-quark decays relies on the correct 
simulation of the shape of the \rel~distribution for all processes.
The simulation was checked with the data. 
For this purpose, an inclusive DIS data sample with at least one
hard jet in the Breit frame was selected, without requiring a muon
in the final state. For tracks passing the same selection criteria as required
for the muon, the \rel~distribution was calculated.
\Fig{control}a shows the comparison of the shape of the measured 
\rel~distribution with the simulated light- and charm-quark contribution.
The shape is reasonably well described.

\Fig{control}b shows the measured \rel~distribution 
for muon candidates  compared to the MC simulation.
The MC simulation contains the background
processes from light and charm quarks and the contribution from $b$ quarks.
The distributions are peaked at low \rel~values, where the decays 
of hadrons containing charm and light quarks dominate. 
At higher \rel~values, the measured distribution falls less 
steeply than that expected for light-quark and charm contributions alone. 
To determine the $b$-quark fraction in the data, 
the contributions from light-plus-charm flavours and beauty in the simulation
 were allowed to vary, and the best fit was 
extracted using a binned maximum-likelihood method. 
The measured fraction of events from $b$ decays, $f_{b}$, 
is $(30.2 \pm 4.1)\%$, where the error is statistical. 
The mixture with the fitted fractions describes the data well.

Figures 1c-f show the comparison between the data and the MC 
simulation with respect to the momentum and the pseudorapidity 
of the muon, as well as the associated jet transverse energy 
in the Breit frame and the pseudorapidity 
of the associated jet measured in the laboratory frame. 
The MC simulation, with the different contributions weighted according to the fractions
found using the fit procedure described above, reproduces the muon and jet kinematics
well.
\section{Systematic uncertainties}
\label{sec-su}
The systematic uncertainties on the measured cross sections were
determined by changing the selection cuts or the analysis procedure
in turn and repeating the extraction of the cross sections. 
The numbers given below refer to the total visible cross section, 
$\sigma_{b\bar{b}}$. For the differential distributions
the systematic uncertainties
were determined bin-by-bin and are included in the figures and in \tab{diff}.
The following systematic studies were carried out:
\begin{itemize}
\item selection cuts and SRTD alignment: 
  variation of the selection cuts by one 
  standard deviation (including the electron energy, $E-p_Z$, 
  $E_{T,\rm{jet}}^{\rm{Breit}}$, $\eta_{\rm{jet}}^{\rm{lab}}$ and 
  SRTD box cut). This led to a systematic deviation of $+9.1 \%$ 
  and $-6.1 \%$ with respect to the nominal value, where the biggest 
  uncertainties were introduced by the widened $\eta_{\rm{jet}}^{\rm{lab}}$ 
  cut and the increased $E_{T,\rm{jet}}^{\rm{Breit}}$ cut. 
  The relative alignment between the RCAL and the SRTD detector
  is known to a precision of $\pm 1$ mm~\cite{thesis:goebel:2001}. 
  The related systematic uncertainty was estimated by shifting the 
  reconstructed SRTD hit position by $\pm 2$ mm in both coordinates
  and was $+0.5 \%$ and $-1.3 \%$, respectively;
\item energy scale: 
  the effect of the uncertainty in the absolute CAL energy scale 
  of $\pm 2\%$ for hadrons  and of 
  $\pm 1\%$ for electrons was $+3.3 \%$ and $-0.3 \%$;
\item extraction of $b$ decays: 
  the uncertainties related to the signal extraction
  were estimated by doubling and halving the charm contribution. This leads to 
  a systematic uncertainty of $+5.7 \%$ and $-3.5\%$, respectively. 
  The uncertainty obtained by reweighting 
  the light-plus-charm quark \rel~distribution with the one extracted from the 
  data as described in Section \ref{sec-bf} is within this uncertainty;
\item muon reconstruction efficiency:
  the effect of the uncertainty on the muon reconstruction efficiency for the barrel and 
  rear regions of the muon detectors was $+8.9 \%$ and $-7.8 \%$;

\item model dependence of acceptance corrections:
  to evaluate the systematic uncertainties on the detector corrections, 
  the results obtained with {\sc Rapgap} were compared with other MC models: {\sc Cascade};
  {\sc Rapgap} with the Colour Dipole Model~\cite{*np:b306:746,*zfp:c43:625};
  and {\sc Rapgap} with the Peterson fragmentation function~\cite{pr:d27:105}.
  Two different values of the $\epsilon$ parameter of the Peterson
  fragmentation function were used, namely $\epsilon=0.0055$ and 0.0041
  as recently determined in $e^+e^-$ collisions by the SLD and OPAL 
  collaborations, respectively~\cite{*pr:d65:092006,*epj:c29:463}.
  The corresponding systematic uncertainty was defined as the maximal
  deviation with respect to the reference sample and was $+2.2 \%$.
                                
\end{itemize}

These systematic uncertainties were added in quadrature separately for the positive and 
negative variations to determine the overall systematic uncertainty. These estimates were 
also made in each bin in which the differential cross sections were measured.
The uncertainty associated with the luminosity measurement for the 
1999-2000 data-taking periods used in this analysis was $\pm 2.2\%$. 
This introduces an overall normalisation uncertainty 
on each measured cross section, which is correlated between all data points.
This is added in quadrature to the other systematic uncertainties on the
 total visible cross section, but is not included in the figures
 or tables of the differential cross section measurements.

%% file: txt-res.tex
%
\section{Results}
\label{sec-res}

The total visible cross section, $\sigma_{b\bbar},$ was determined 
in the  kinematic range $Q^2>2\gev^2$, $0.05<y<0.7$ with at least one hadron-level 
jet in the Breit frame with $E_{T,\rm{jet}}^{\rm{Breit}}>6\gev$ and 
$-2 < \eta_{\rm{jet}}^{\rm{lab}} < 2.5$ and with a muon 
fulfilling the following conditions: 
$-0.9 < \eta^{\mu} < 1.3$ and $p_T^{\mu}>2\gev$ or
$-1.6 < \eta^{\mu} < -0.9$ and $p^{\mu}>2\gev$. 
The jets were defined by applying the $k_T$ algorithm to stable hadrons; 
weakly decaying {\it B} hadrons are considered unstable. 
The muons coming from direct and indirect $b$ decays are matched to any jet in the event. 
The measured cross section is
$$\sigma_{b\bbar}(ep\rightarrow e~b~\bbar~X 
\rightarrow e~{\rm jet}~\mu~X) = 
40.9 \pm 5.7\;{\rm (stat.)} ^{+6.0}_{-4.4} {\rm (syst.)} \pb.$$

This measurement has been  corrected for electroweak radiative effects
using {\sc Heracles}.
The NLO QCD prediction with hadronisation corrections 
is $20.6 ^{+3.1} _{-2.2} \; \rm{pb}$, which is
about 2.5 standard deviations lower than the measured total cross section.
The {\sc Cascade} MC program gives $\sigma_{b\bbar} = 28 \pb$ and
{\sc Rapgap} gives $\sigma_{b\bbar} = 14 \pb$.

The differential cross sections were calculated in the same restricted 
kinematic range as the total cross section by repeating the fit of the 
\rel~distribution and evaluating the electroweak radiative corrections
in each bin. The results are summarised in \tab{diff}.

Figures~\ref{fig-q2x}a and~\ref{fig-q2x}b show the differential cross sections as functions 
of $Q^2$ and $x$, respectively,  compared to the NLO QCD calculation. 
The NLO QCD predictions generally
agree with the data; in the lowest $Q^2$ and lowest $x$ bins,
the data are about two standard deviations higher.
Figures~\ref{fig-q2x}c and~\ref{fig-q2x}d show the same differential cross sections compared
with the {\sc Rapgap} and {\sc Cascade} MC simulations. {\sc Cascade} agrees with the
data except for the lowest $Q^2$ and lowest $x$ bin. {\sc Rapgap} is well below the
data in all bins, but it reproduces the shapes of the data distributions.

Figures~\ref{fig-pteta}a and~\ref{fig-pteta}b show the differential cross sections 
as functions of the transverse momentum, $p_T^{\mu}$, and pseudorapidity, 
$\eta^{\mu}$, of the muon, compared to the NLO QCD calculation.
They generally agree with the data; in the lowest  $p_T^{\mu}$ bin and the high $\eta^{\mu}$
 bin, the NLO QCD prediction is about two standard deviations below the data.
Figures~\ref{fig-pteta}c and~\ref{fig-pteta}d show the same differential distribution 
compared with {\sc Cascade} and {\sc Rapgap}. {\sc Cascade} describes the measured 
cross sections well except for the lowest $p_T^{\mu}$ bin, while {\sc Rapgap} lies below the data.

\Fig{etjet}a shows the differential cross section as a
function of $E_{T,\rm{jet}}^{\rm{Breit}}$
of the leading jet compared to the NLO QCD calculation. 
The NLO QCD prediction agrees with the data reasonably well, 
though it is systematically below. 
For the highest $E_{T,\rm{jet}}^{\rm{Breit}}$ bin the difference is about
two standard deviations.
\Fig{etjet}b shows the same differential distribution compared with {\sc Cascade}
and {\sc Rapgap}. For all $E_{T,\rm{jet}}^{\rm{Breit}}$ values, {\sc Cascade} reproduces
the measured cross section reasonably well while {\sc Rapgap} lies below the data.

%% file: txt-con.tex
\section{Conclusions}
\label{sec-con}
The production of $b$ quarks in the deep inelastic scattering
process $ep \rightarrow e~\mu~{\rm jet}~X$ has been measured with the 
ZEUS detector at HERA. 
The NLO QCD prediction for the visible 
cross section lies about 2.5 standard deviations below 
the measured value.

Single differential cross sections as functions of the photon virtuality, 
$Q^2$, the Bjorken scaling variable, $x$,
the transverse momentum and pseudorapidity of the muon as well
as the transverse energy of the leading jet in the Breit frame
have been measured.
The {\sc Cascade} MC program, implementing the CCFM QCD evolution
equations, gives a good description of the measured cross sections.
It is, however, below the data for low values of 
the transverse momenta, low $Q^2$ and low values of $x$.
{\sc Rapgap} is well below the data for all measured cross sections.
The differential cross sections are in
general consistent with the NLO QCD predictions; however at low values
of $Q^2$, Bjorken $x$, and muon transverse momentum, and high values of
jet transverse energy and muon pseudorapidity, the prediction is about
two standard deviations below the data

In summary, $b$-quark production in DIS has been measured for the first time
and has been shown to be consistent with NLO QCD calculations.

%% file: txt-ack.tex
\section*{Acknowledgements}
\label{sec-ack}
We thank the DESY directorate for their strong support and encouragement.
The special efforts of the HERA group are gratefully acknowledged.
We are grateful for the support of the DESY computing and network services.
The design, construction and installation of the ZEUS detector have been
made possible by the ingenuity and effort of many people who are not listed 
as authors. We thank B.W.~Harris and J.~Smith for providing the NLO code.
We also thank H.~Jung and M.~Cacciari for useful discussions.

%% file: paper-ref.tex
{
\def\bibname{\Large\bf References}
\def\refname{\Large\bf References}
\pagestyle{plain}
\ifzeusbst
  \bibliographystyle{./BiBTeX/bst/l4z_default}
\fi
\ifzdrftbst
  \bibliographystyle{./BiBTeX/bst/l4z_draft}
\fi
\ifzbstepj
  \bibliographystyle{./BiBTeX/bst/l4z_epj}
\fi
\ifzbstnp
  \bibliographystyle{./BiBTeX/bst/l4z_np}
\fi
\ifzbstpl
  \bibliographystyle{./BiBTeX/bst/l4z_pl}
\fi
{\raggedright
\bibliography{./BiBTeX/user/syn.bib,%
              ./BiBTeX/bib/l4z_articles.bib,%
              ./BiBTeX/bib/l4z_kkarticles.bib,%
              ./BiBTeX/bib/l4z_books.bib,%
              ./BiBTeX/bib/l4z_kkbooks.bib,%
              ./BiBTeX/bib/l4z_conferences.bib,%
              ./BiBTeX/bib/l4z_h1.bib,%
              ./BiBTeX/bib/l4z_misc.bib,%
              ./BiBTeX/bib/l4z_old.bib,%
              ./BiBTeX/bib/l4z_preprints.bib,%
              ./BiBTeX/bib/l4z_kkpreprints.bib,%
              ./BiBTeX/bib/l4z_replaced.bib,%
              ./BiBTeX/bib/l4z_temporary.bib,%
              ./BiBTeX/bib/l4z_kktemporary.bib,%
              ./BiBTeX/bib/l4z_zeus.bib}}
}
\vfill\eject

%% file: paper-tab.tex
\begin{table}[p]
\begin{center}
\begin{tabular}{|c|ccc|}
\hline 
$Q^2$ range & $d\sigma/dQ^2$  &                 &                 \\
(\gev$^2$)  & (pb/$\gev^2$)   & $\Delta_{stat}$ & $\Delta_{syst}$ \\
\hline \hline
$2,10$    &   2.63  & $\pm 0.56$  & $^{+0.53}_{-0.46}$ \\
$10,40$   &   0.36  & $\pm 0.10$  & $^{+0.06}_{-0.05}$ \\
$40,1000$ &   0.010 & $\pm 0.002$ & $^{+0.002}_{-0.002}$ \\
\hline \hline 
$\log_{10}(x)$ range  & $d\sigma/dx$   & & \\
                      &      (pb)      & $\Delta_{stat}$ & $\Delta_{syst}$ \\
\hline \hline
$ -4.5, -3.5$  & 20.9 & $\pm 4.4$ & $^{+3.2}_{-3.4}$ \\
$ -3.5, -2.9$  & 17.2 & $\pm 4.7$ & $^{+2.3}_{-2.5}$ \\
$ -2.9, -1.0 $ &  5.3 & $\pm 1.3$ & $^{+0.9}_{-1.0}$ \\
\hline \hline 
$p_T^{\mu}$ range & $d\sigma/dp_T^{\mu}$ & & \\
      (\gev)      &     (pb/\gev)        & $\Delta_{stat}$ & $\Delta_{syst}$ \\
\hline \hline
$2,3$   & 30.5 & $\pm 7.6$    & $^{+6.3}_{-4.2}$ \\
$3,4$   &  9.7 & $\pm 2.6$    & $^{+1.9}_{-1.8}$ \\
$4,15$   & 0.59 & $\pm 0.13$   & $^{+0.11}_{-0.13}$ \\
\hline \hline 
$\eta^{\mu}$ range & $d\sigma/d\eta^{\mu}$  & & \\
                   &        (pb)            & $\Delta_{stat}$ & $\Delta_{syst}$ \\
\hline \hline
$-1.6, -0.15$  &   9.1 & $\pm 2.2$  & $^{+1.9}_{-1.5}$ \\
$-0.15, 0.45$  &  14.2 & $\pm 3.6$  & $^{+3.0}_{-3.0}$ \\
$0.45,  1.3 $  &  19.8 & $\pm 4.1$  & $^{+3.8}_{-3.1}$ \\
\hline \hline 
$E_{T,\rm{jet}}^{\rm{Breit}}$ range & $d\sigma/dE_{T,\rm{jet}}^{\rm{Breit}}$ & & \\
 (\gev)                            &  (pb/\gev)                              &  $\Delta_{stat}$ & $\Delta_{syst}$ \\
\hline \hline
$6, 10$ & 5.7  & $\pm 1.4$  & $^{+1.4}_{-1.3}$ \\
$10,13$ & 3.4  & $\pm 0.8$  & $^{+0.5}_{-0.4}$ \\
$13,36$ & 0.40 & $\pm 0.08$ & $^{+0.05}_{-0.05}$ \\
\hline
\end{tabular}
\vspace{.8cm}
\caption{Single differential $b$-quark cross sections as functions of
$Q^2$, the Bjorken-$x$ variable, the muon transverse momentum, $p_T^\mu$,
the muon pseudorapidity, $\eta^\mu$, and the transverse energy of the
leading jet in the Breit frame, $E_{T,\rm{jet}}^{\rm{Breit}}$. The 
statistical and systematic uncertainties are shown separately.}
\label{tab-diff}
\end{center}
\end{table}

%% file: paper-fig.tex
%

\begin{figure}[p]
%
%
\begin{center}
\epsfig{file=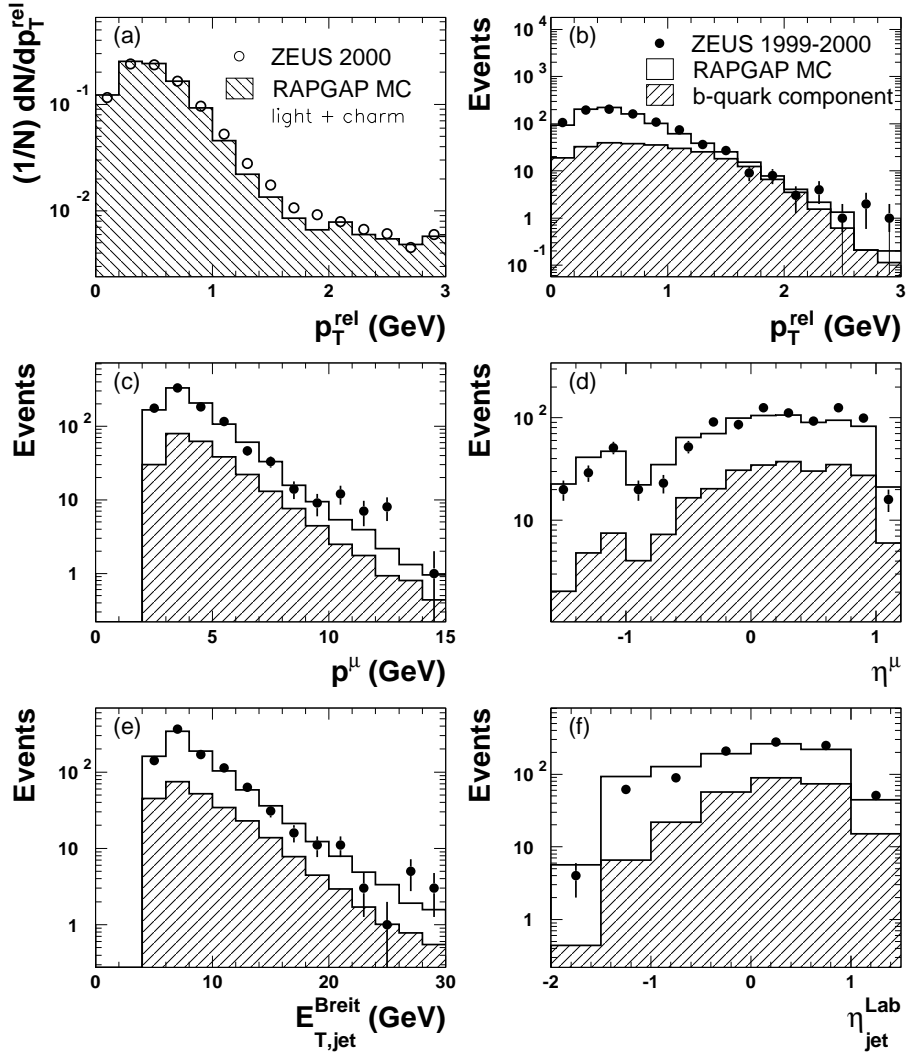, bbllx=0, bblly=0, bburx=567, bbury=650 , width=13cm}
\vspace*{-1.5cm}
\caption{
(a) \rel~distribution measured for unidentified tracks 
in an inclusive DIS sample compared with the {\sc Rapgap} MC simulation 
 ({\it see text}).
Data (dots) and the {\sc Rapgap} MC (solid line) distributions after
the final event selection for: (b) the measured \rel~distribution;
(c) muon momentum; (d) muon pseudorapidity; 
(e) transverse energy in the Breit frame; and (f) pseudorapidity in the 
laboratory frame of the associated jet. The solid line represents all MC 
contributions while the hatched histograms show the contribution 
from $b$ quarks according to the percentage given by the fit (see \Sect{res}).
The error bars are statistical only.
}
\label{fig-control}
\end{center}
\end{figure}
\clearpage
\begin{figure}[p]
%
%
\begin{center}
\epsfig{file=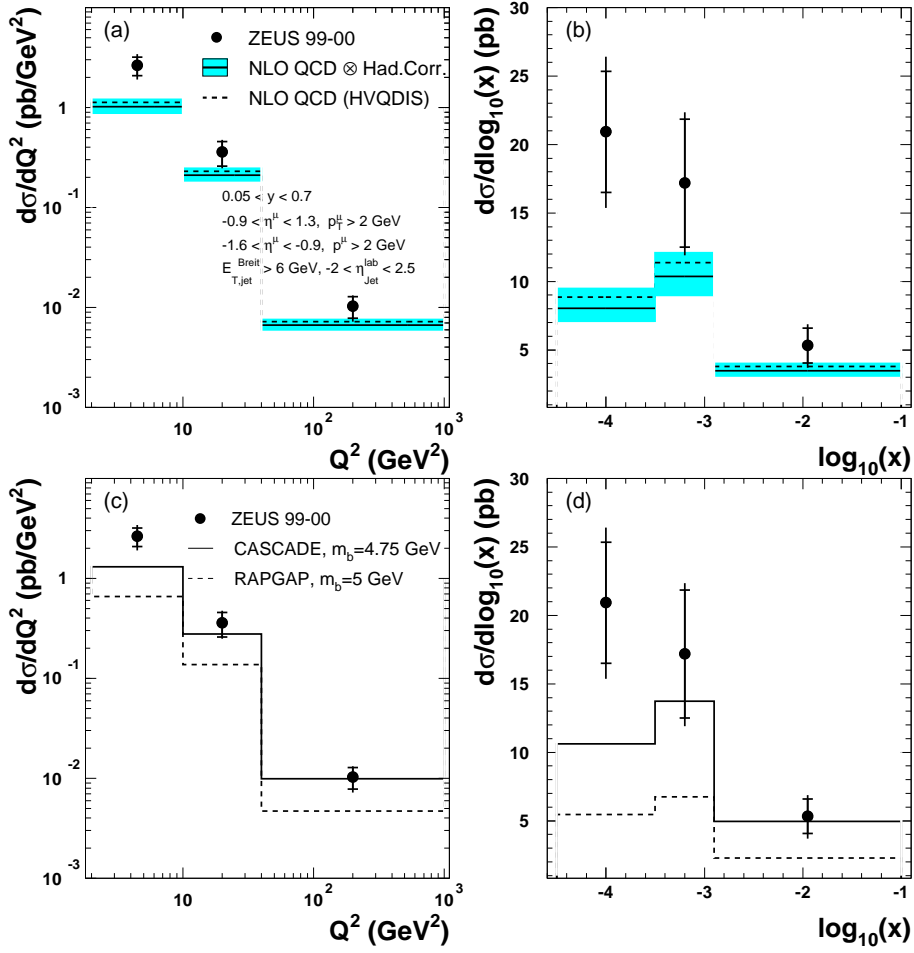, bbllx=0, bblly=0, bburx=567, bbury=650 , width=13cm}
\vspace*{-1.5cm}
\caption{Differential $b$-quark cross section as a function of (a) $Q^2$ and 
(b) Bjorken $x$
for events with at least one jet reconstructed in the Breit frame 
and a muon, compared to the NLO QCD calculations. The error bars on the data 
points correspond to the statistical uncertainty (inner error bars) and 
to the statistical and systematic uncertainties added in quadrature 
(outer error bars). 
The solid line shows the NLO QCD calculations with the hadronisation
corrections and the dashed line the same calculation without the
hadronisation corrections.
The shaded bands show the uncertainty of the 
NLO QCD prediction due to the variation of the renormalisation and 
factorisation scale,  $\mu$, and the $b$-quark mass,  $m_b$.  
Differential $b$-quark cross sections as a function of (c) $Q^2$ and (d) Bjorken $x$, 
compared with the LO QCD MC programs {\sc Cascade} (solid line) and {\sc Rapgap} (dashed
line).}
\label{fig-q2x}
\end{center}
\end{figure}
\clearpage
\begin{figure}[p]
%
%
\begin{center}
\epsfig{file=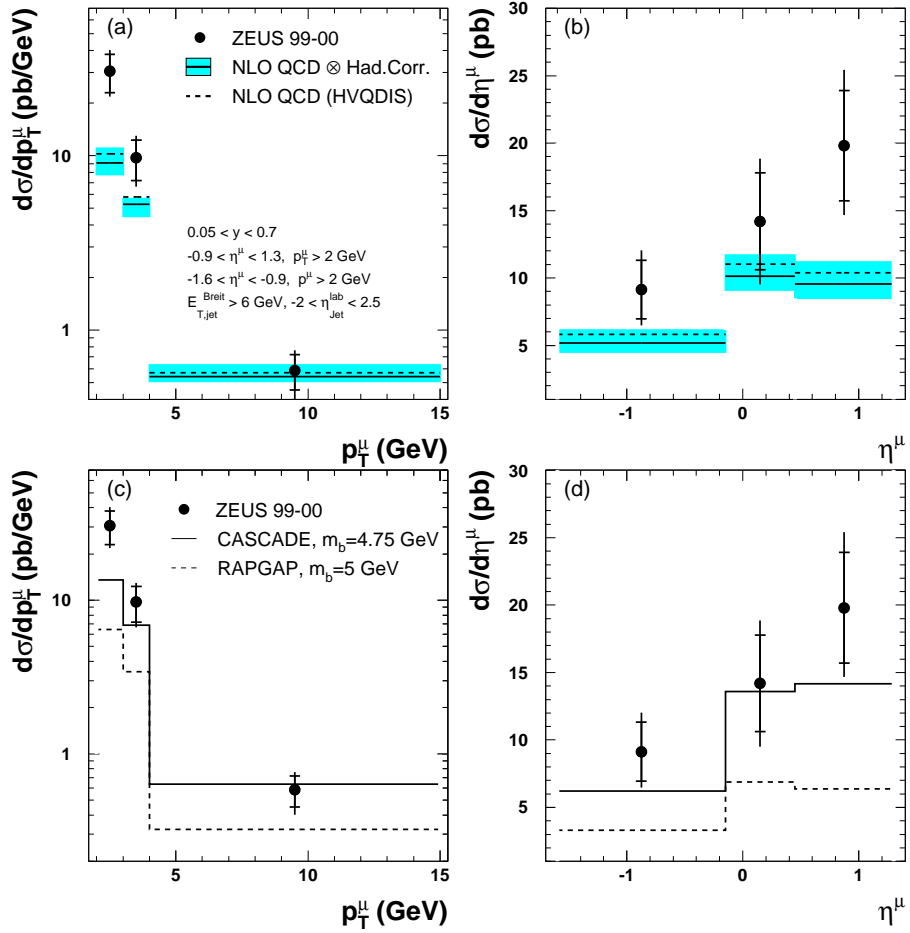, bbllx=0, bblly=0, bburx=567, bbury=650 , width=13cm}
\vspace*{-1.5cm}
\caption{Differential $b$-quark cross section as a function of (a) the 
muon transverse momentum $p_T^{\mu}$ and (b) muon pseudorapidity 
$\eta^{\mu}$ in the laboratory frame, compared to the NLO QCD 
calculations. Other details are as described in the caption to \fig{q2x}.
Differential $b$-quark cross section as a function of (c) $p_T^{\mu}$
and  (d) $\eta^{\mu}$,
compared with LO QCD MC programs {\sc Cascade} (solid line) and {\sc Rapgap} (dashed
line).}
\label{fig-pteta}
\end{center}
\end{figure}
\clearpage
\begin{figure}[p]
%
%
\begin{center}
\epsfig{file=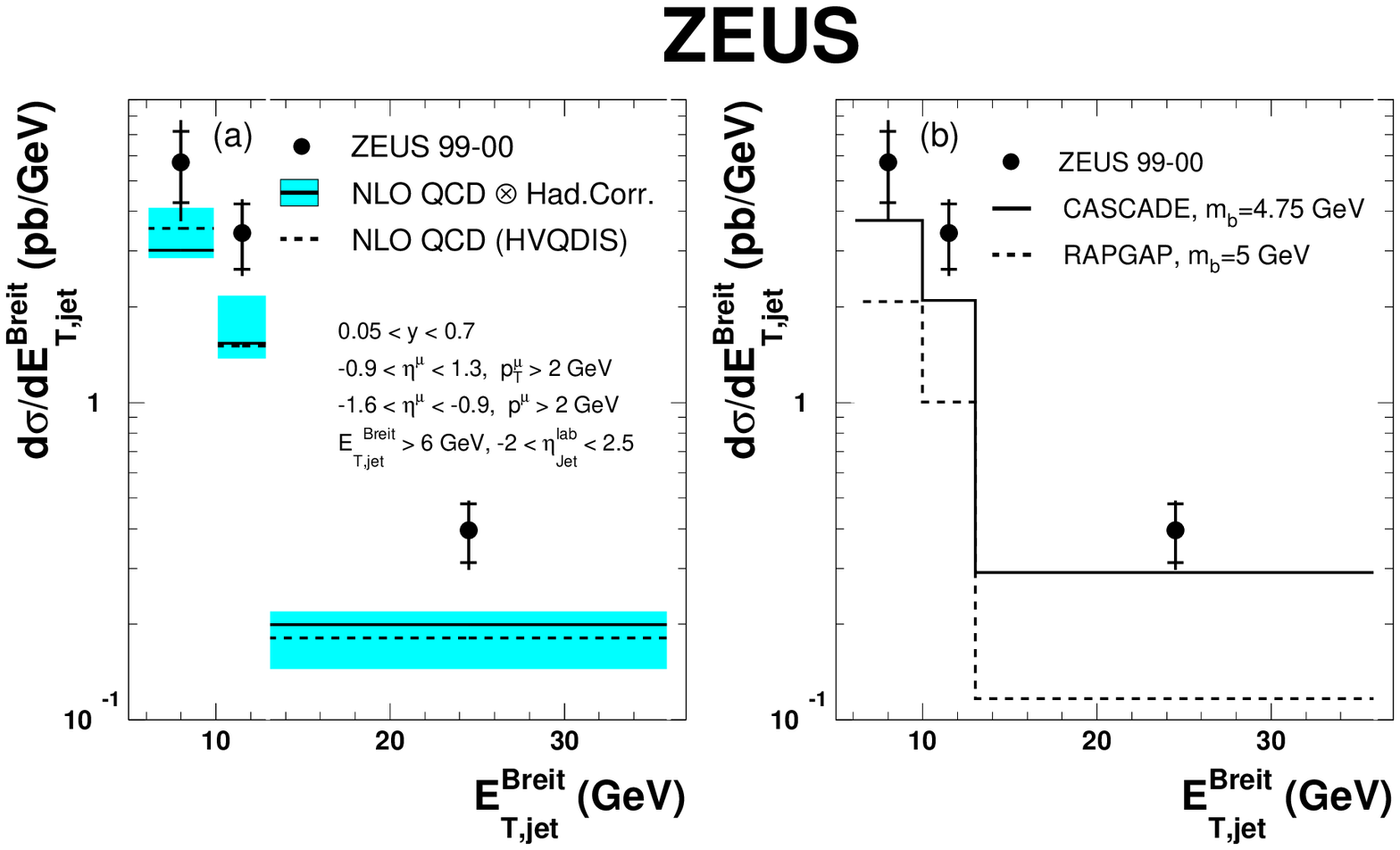, bbllx=0, bblly=0, bburx=567, bbury=650 , width=13cm}
\vspace*{-8.5cm}
\caption{(a) differential $b$-quark cross section as a function of the 
transverse energy of the jet in the Breit frame $E_{T,\rm{jet}}^{\rm{Breit}}$.
The data (dots) are compared to the NLO QCD calculations (a).
Other details are as described in the caption to \fig{q2x}.
(b) differential $b$-quark cross sections as a function of $E_{T,\rm{jet}}^{\rm{Breit}}$
compared with LO QCD MC programs {\sc Cascade} (solid line) and {\sc Rapgap} (dashed
line).}
\label{fig-etjet}
\end{center}
\end{figure}